\documentclass[letterpaper,10pt]{article}

\usepackage[T1]{fontenc}
\usepackage[utf8]{inputenc}

\usepackage[left=1.5cm,right=1.5cm,top=3cm,bottom=1.5cm]{geometry}

\usepackage{cite}
\usepackage{amsmath,amssymb,amsfonts}
\usepackage{algorithmic}
\usepackage{graphicx}
\usepackage{textcomp}
\usepackage{xcolor}

\usepackage{booktabs}
\usepackage{url}
\usepackage{setspace}
\usepackage{titlesec}
\usepackage{caption}
\usepackage{array}
\usepackage{enumitem}
\usepackage{ragged2e}
\usepackage{fancyhdr}

\usepackage{helvet}

\definecolor{headergray}{RGB}{105,105,105}
\definecolor{linegray}{RGB}{200,200,200}

\newcommand{\mycustomheader}{%
    \begin{minipage}[c]{2cm}
    \end{minipage}%
    \hspace{0.2cm}%
    \begin{minipage}[c]{7cm}
        \centering
        \color{headergray}
    \end{minipage}

    \vspace{6pt}
    {\color{linegray}\hrule height 0.5pt}
    \vspace{6pt}
}

\fancypagestyle{plain}{
  \fancyhf{}
  \fancyhead[C]{\mycustomheader}
  
}

\titleformat{\section}
  {\fontsize{10pt}{12pt}\bfseries\MakeUppercase}
  {}
  {0pt}
  {}
\titlespacing*{\section}{0pt}{0pt}{12pt}

\titleformat{\subsection}
  {\fontsize{9pt}{10pt}\bfseries\itshape}
  {}
  {0pt}
  {}
\titlespacing*{\subsection}{0pt}{0pt}{12pt}

\begin{document}

\begin{center}
{\fontsize{12pt}{14pt}\bfseries\MakeUppercase
Impact of Physics-Informed Features on Neural Network Complexity for Li-ion Battery Voltage Prediction in Electric Vertical Takeoff and Landing Aircrafts
\par}
\vspace{18pt}

{\fontsize{9pt}{11pt}\itshape
\textsuperscript{1,2*} Eymen Ipek,
\textsuperscript{2} Assoc. Prof. Mario Hirz,
\par}

{\fontsize{9pt}{11pt}
\textsuperscript{1} Virtual Vehicle Research GmbH, Graz, Austria
\par
\textsuperscript{2} Institute of Automotive Engineering, Graz University of Technology, Graz, Austria\par
}

\vspace{6pt}
{\fontsize{9pt}{11pt}
*Corresponding author e-mail: eymen.ipek@v2c2.at
}
\end{center}
\section*{ABSTRACT}
\vspace{-6pt}
\justifying
The electrification of vertical takeoff and landing aircraft demands high-fidelity battery management systems capable of predicting voltage response under aggressive power dynamics. While data-driven models offer high accuracy, they often require complex architectures and extensive training data. Conversely, equivalent circuit models (ECMs), such as the second-order model, offer physical interpretability but struggle with high $C$-rate non-linearities. This paper investigates the impact of integrating physics-based information into data-driven surrogate models. Specifically, we evaluate whether physics-informed features allow for the simplification of neural network architectures without compromising accuracy. Using the open-source electric vertical takeoff and landing (eVTOL) battery dataset, we compare pure data-driven models against physics-informed data models. Results demonstrate that physics-informed models achieve comparable accuracy to complex pure data-driven models while using up to 75\% fewer trainable parameters, significantly reducing computational overhead for potential on-board deployment.

\vspace{6pt}
\noindent\textbf{Keywords:} Physics-informed machine learning, battery modeling, eVTOL, surrogate models, edge AI

\vspace{12pt}
\section{Introduction}
\vspace{-6pt}
Lithium-ion batteries are the preferred energy storage solution for electric aviation due to their high energy density. However, electric vertical takeoff and landing (eVTOL) mission profiles are characterized by distinct high-power pulses during takeoff and landing, which induce rapid voltage drops and thermal stress \cite{bills2023}. Given the complexity of environmental conditions and the necessity of precise flight planning, eVTOLs significantly benefit from robust future state prediction. Accurate voltage prediction is critical for State-of-Charge (SOC) and State-of-Power (SOP) estimation to ensure safety margins during these critical flight phases.The current modeling landscape faces distinct trade-offs between fidelity and computational efficiency. Physics-based models, such as the Pseudo-Two-Dimensional (P2D) model, describe internal electrochemical states governed by partial differential equations (PDEs). While highly accurate, they are computationally prohibitive for real-time control \cite{newman2004}. Consequently, traditional Battery Management Systems (BMS) rely on Equivalent Circuit Models (ECMs), such as the Thevenin or 2RC models. These models balance complexity and speed by approximating dynamics of electrical components ($R, C$) \cite{hu2012}. However, ECMs often underperform under the highly dynamic current profiles, which are typical eVTOLs due to parameter variability and unmodeled electrochemical non-linearities.To address these non-linearities, Deep Neural Networks (DNNs) have emerged as a powerful alternative \cite{ng2020}. Data-driven methods can capture complex behaviors that escape traditional ECMs. Yet, purely data-driven models act as "black boxes" and often require heavy computational resources, both in memory and inference time. This poses significant challenges for embedded BMS hardware, where edge AI capabilities are constrained. Furthermore, pure data-driven models suffer from poor extrapolation issues when operating outside their training domain \cite{ng2020}. A promising direction represents the Physics-Informed Machine Learning (PIML) approach, which aims to merge the interpretability of physics with the flexibility of learning. Integrating physics with machine learning can significantly enhance model fidelity of FNNs, as demonstrated in recent work by Tu et al. \cite{tu2023} for Single Particle Models (SPM) and Nonlinear Double Capacitor (NDC) models. 

Building on this foundation, the present paper proposes a PIML approach where a 2RC model guides the learning process of a neural network. The primary objective is to quantify how this physics injection influences the required size of the neural network. We hypothesize that this physics integration enables drastic model compression without compromising fidelity. Indeed, our results demonstrate that a minimal single-layer Physics-Informed Neural Network (PINN) creates a superior surrogate model, outperforming significantly deeper Feedforward Neural Networks (FNN) by reducing both Root Mean Square Error (RMSE) and peak transient errors by approximately 50\%.

\vspace{12pt}

\section{Methodology}
\vspace{-6pt}
To validate the proposed modeling framework, we utilize the eVTOL battery dataset, as published by Bills et al. \cite{bills2023}. This dataset captures high-fidelity cycling data of lithium-ion batteries under dynamic mission profiles representative for eVTOL applications. For this study, five distinct cells are selected to evaluate model generalization across varying operating conditions. The dataset is partitioned into a training set consisting of four cells and a testing set consisting of a single unseen cell.
The training set comprises cells VAH05, VAH10, VAH12, and VAH26. To ensure that the models capture dynamics across the entire battery lifespan, training data is specifically sampled from the 1\textsuperscript{st}, 50\textsuperscript{th}, and 1000\textsuperscript{th} cycles. These cells expose the network to a diverse range of power demands and environmental conditions \cite{bills2023}:
\begin{itemize}
    \item \textbf{VAH05:} Operate with a power reduction of 10\% during takeoff, cruise, and landing phases.
    \item \textbf{VAH10:} Cycles within a thermal chamber maintained at 30°C.
    \item \textbf{VAH12:} Is subjected to a mission profile with a cruise time ($t_c$) of 400 seconds.
    \item \textbf{VAH26:} Is subjected to a mission profile with an extended cruise time of 600 seconds.
\end{itemize}

For model validation, cell \textbf{VAH11} is isolated as the testing target. This cell operates under a distinct condition characterized by a 20\% power reduction across takeoff, cruise and landing phases. All model evaluations presented in this work conducted on the 600\textsuperscript{th} cycle of VAH11 to assess performance at a mature state of aging. Two distinct surrogate modeling approaches are developed such as FNN and PINN. 

The FNN implements a purely data-driven regression where the predicted voltage $V_{\text{pred}}$ is a direct function $f_\theta(x)$ of the measured inputs. The model relies solely on measured or derived state variables. The learning objective minimizes the Mean Squared Error (MSE) between the predicted and normalized measured voltage:

\begin{equation}
\hat{\theta} \in \arg \min_{\theta} \frac{1}{N} \sum_{i=1}^{N} \left\| \tilde{V}_i - f_{\theta}(\tilde{x}_i) \right\|_2^2
\end{equation}


\begin{itemize}
\item $x_i$: The input feature vector represents the raw operating conditions and consists of step time $\Delta t$, current $I$, $\text{SoC}$, temperature $T$ and cycle number $N$:
\begin{itemize}
\item $x = [\Delta t, I, \text{SOC}, T, N]$
\end{itemize}
\item $y_i$: The target is the terminal voltage $\tilde{V}_i$, derived from the measured voltage $V_i$.
\item $f_\theta$: The function represents the feed-forward network, which maps the input vector $x$ to the predicted voltage by using ReLU-activated hidden layers and a linear output layer.
\end{itemize}

The PINN framework proposed in this study is a hybrid residual model. Rather than forcing the network to relearn well-understood electrochemical dynamics, we embed an equivalent circuit prior. A second-order RC (2RC) model computes a physics-based voltage guess via discrete-time state updates:
\begin{equation}
V_{\text{phy},i} = \text{OCV}_i - I_i R_0 - V_{RC1,i} - V_{RC2,i}
\end{equation}

The internal polarization states $V_{RCj}$ (for $j=1, 2$) are updated at each timestep $\Delta t$ according to:

\begin{equation}
V_{RCj,i} = e^{-\Delta t_i / \tau_j} V_{RCj,i-1} + R_j \left( 1 - e^{-\Delta t_i / \tau_j} \right) I_i
\end{equation}

The neural network is then tasked only with learning the residual correction term $\Delta V_\theta$, such that the final voltage prediction becomes $V_{\text{pred}} = V_{\text{phy}} + \Delta V_\theta(x)$. This formulation anchors the prediction to physically meaningful dynamics, reducing the hypothesis space the network must search. The corresponding learning objective is defined as:

\begin{equation}
\hat{\theta} \in \arg \min_{\theta} \frac{1}{N} \sum_{i=1}^{N} \left\| V_i - \left( V_{\text{phy},i} + \Delta V_{\theta}(\tilde{x}_i) \right) \right\|_2^2
\end{equation}

\begin{itemize}
\item $x_i$: The input feature vector is augmented with physics-derived terms, thermal conditions and aging factors to facilitate residual learning. It includes current $I$, state-of-charge $\text{SOC}$, temperature $T$, cycle number $N$, equivalent circuit states ($V_{RC1}, V_{RC2}$) and dynamic features:
\begin{itemize}
\item $x = [\Delta t, I, \text{SOC}, T, N, \text{OCV}, V_{RC1}, V_{RC2}, V_{phy}]$
\end{itemize}
\item $y_i$: The target is the measured terminal voltage $V$, and the hybrid model predicts a residual correction to the physics guess:
\begin{itemize}
\item $V_{\text{pred}} = V_{phy} + \Delta V_{\theta}(x)$
\end{itemize}
\end{itemize}

\vspace{12pt}

\section{Results and Discussion}
\begin{figure}[h]
    \centering
    \includegraphics[width=0.8\linewidth]{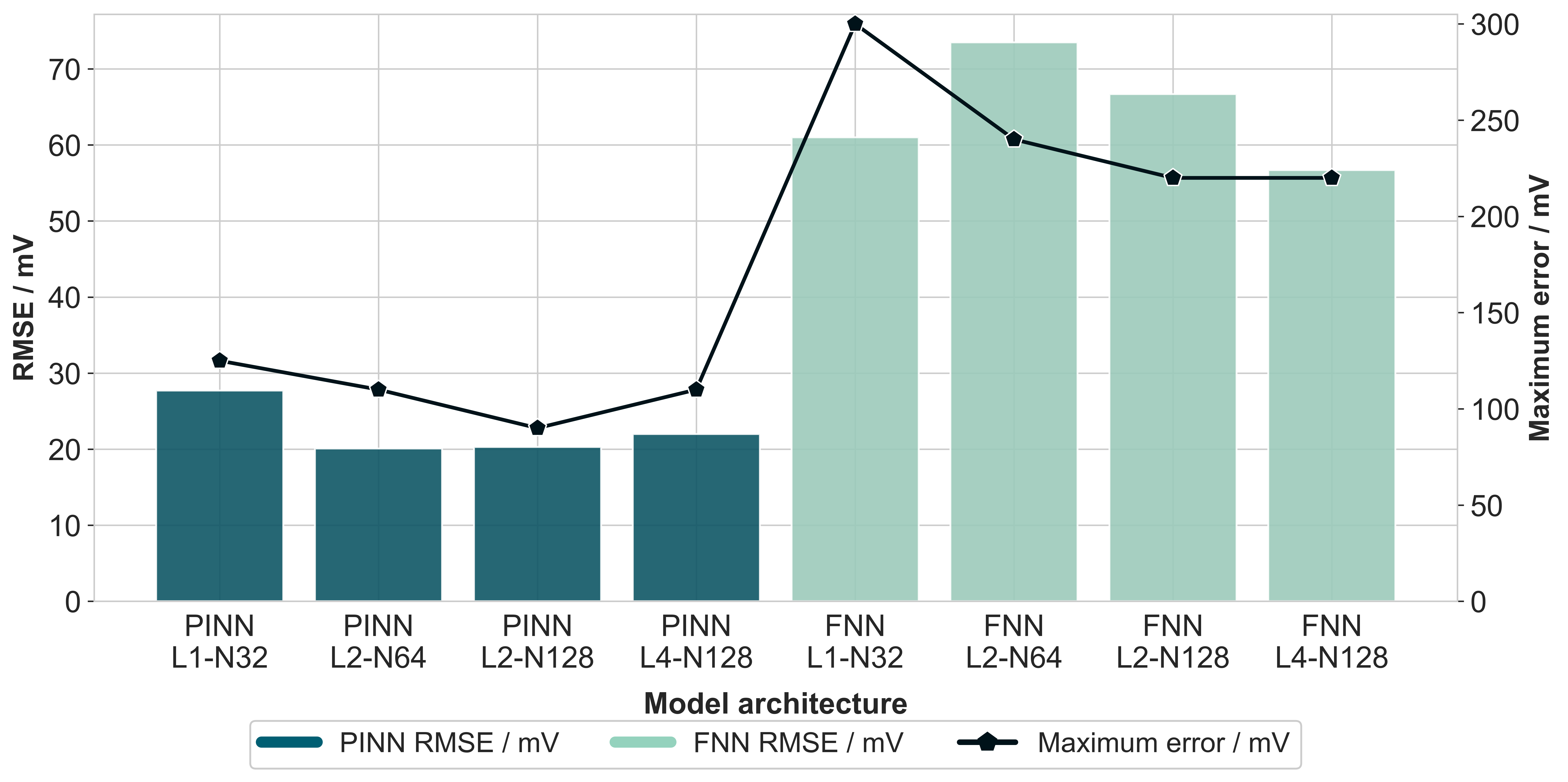}
    \caption{Performance comparison of PINN versus FNN architectures. The PINN architectures consistently achieve significantly lower RMSE and maximum error compared to their FNN counterparts, even when the PINN is much simpler than the most complex FNN tested.}
    \label{fig:performance_comparison}
\end{figure}

\begin{figure}[h]
    \centering
    \includegraphics[width=0.7\linewidth]{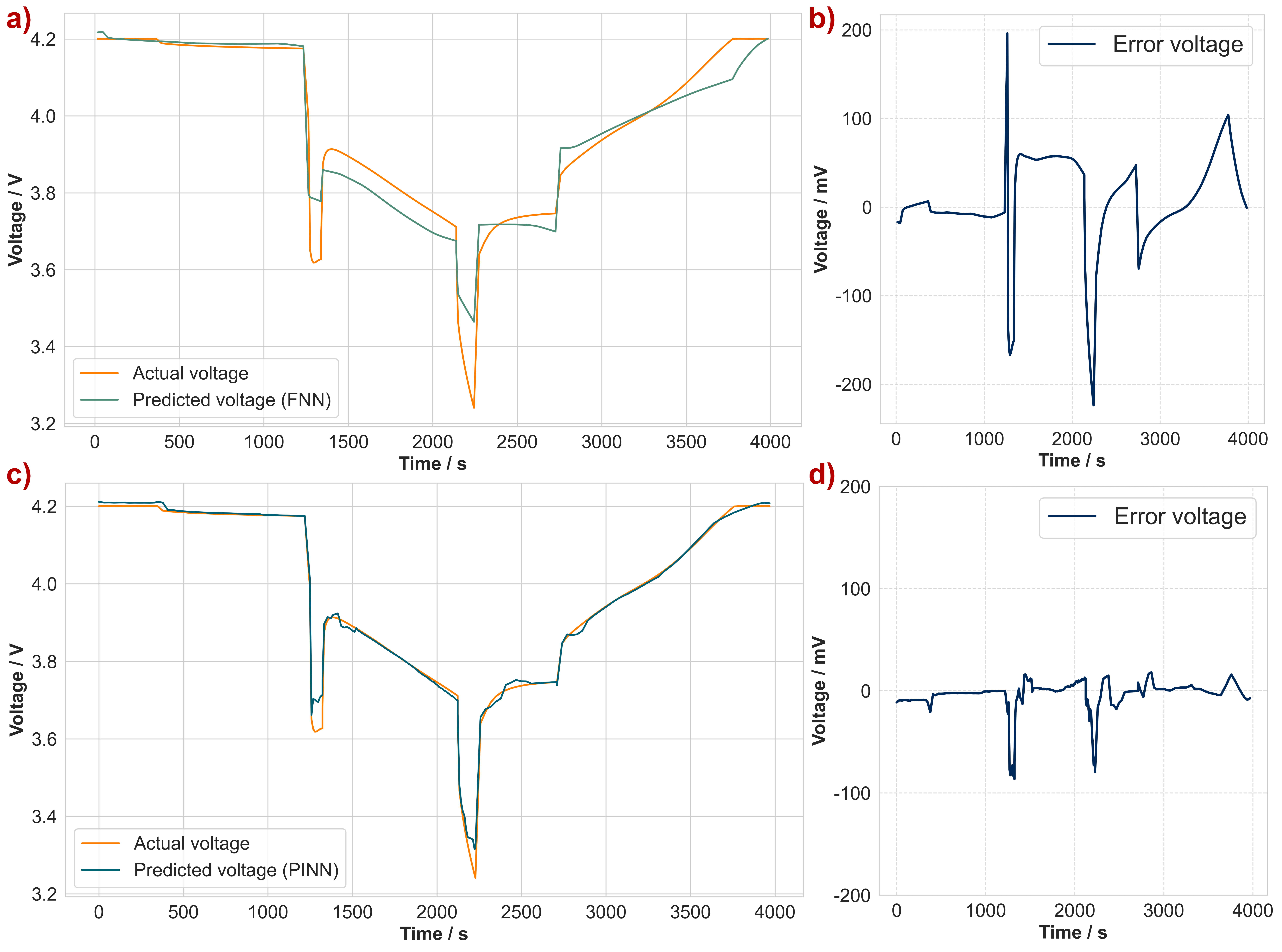}
    \caption{Voltage prediction and error analysis under dynamic eVTOL profiles. (a) the actual voltage versus the FNN (2 hidden layers, 128 neurons each) predicted voltage, (b) the error voltage for the FNN, (c) the actual voltage versus the PINN (2 hidden layers, 128 neurons each) predicted voltage (d) the error voltage for the PINN.}
    \label{fig:Comparison_plots}
\end{figure}

\vspace{-6pt}
We trained multiple configurations varying hidden layers as 1, 2, 4 and neurons as 32, 64, 128. We compared MAE, Root Mean Squared Error (RMSE), $R^2$ Score, and computational timing as shown in Figure~\ref{fig:performance_comparison}, Figure~\ref{fig:Comparison_plots} and summarized in Table~\ref{neural_network_metrics}.

The quantitative results demonstrate a substantial performance advantage for the physics-informed approach; a minimal single-layer PINN (L1-N32) achieved a Root Mean Square Error (RMSE) of 27.7 mV and an $R^{2}$ score of 98.5\%, whereas the equivalent FNN (L1-N32) resulted a significantly higher RMSE of 61.0 mV and a maximum error of 300 mV. Furthermore, while increasing the complexity of the FNN to four layers (L4-N128) improved its RMSE to 39.2 mV, it still failed to surpass the accuracy of the simplest PINN, and the optimal PINN configuration (L2-N64) achieved the lowest RMSE of 20.1 mV with an $R^{2}$ of 99.2\%. Overall, the physics-informed models demonstrated the ability to reduce both RMSE and peak transient errors by approximately 50\% while utilizing up to 75\% fewer trainable parameters than complex pure data-driven models.

\begin{table}[h]
\centering
\begin{tabular}{lcccccccc}
\toprule
\textbf{Model} & \textbf{Hidden layers} & \textbf{Number of neurons} &
\textbf{Maximum error / mV} & \textbf{MAE / mV} &
\textbf{RMSE / mV} & \textbf{$R^2$ score / \%} \\
\midrule
FNN & 1 & 32 & 300 & 51.408 & 61.0 & 92.8 \\
FNN & 2 & 64 & 240 & 55.128 & 73.5 & 89.6 \\
FNN & 2 & 128 & 220 & 48.546 & 66.7 & 91.4 \\
FNN & 4 & 128 & 220 & 39.2 & 56.7 & 93.8 \\
\midrule
PINN & 1 & 32 & 125 & 15.889 & 27.7 & 98.5 \\
{\bfseries PINN} & {\bfseries 2} & {\bfseries 64} & {\bfseries 110} & {\bfseries 9.652} & {\bfseries 20.1} & {\bfseries 99.2} \\
{\bfseries PINN} & {\bfseries 2} & {\bfseries 128} & {\bfseries 90} & {\bfseries 10.109} & {\bfseries 20.3} & {\bfseries 99.2} \\
PINN & 4 & 128 & 110 & 8.919 & 22.0 & 99.1 \\
\bottomrule
\end{tabular}
\caption{FNN and PINN performance metrics summary.}
\label{neural_network_metrics}
\end{table}

\section{CONCLUSION}
\vspace{-6pt}
This paper investigates the impact of integrating physics-based information into data-driven surrogate models to address the challenges of voltage prediction in eVTOL applications. By using a hybrid approach where a second-order ECM guides the learning process, it is confirmed that physics integration enables drastic model compression without compromising fidelity. Therefore, a minimal PINN can act as a superior surrogate model, outperforming significantly deeper FNN and offering a computationally efficient solution suitable for the resource-constrained edge AI environments found in onboard BMS and flight controllers.
\vspace{12pt}
\section*{ACKNOWLEDGEMENT}
\vspace{-6pt}
This work was funded by the program “Industrienahe Dissertation 2024” of the Austrian Federal Ministry for Innovation, Mobility and Infrastructure (BMIMI). 
The publication was written at Virtual Vehicle Research GmbH in Graz and partially funded within the COMET K2 Competence Centers for Excellent Technologies from the Austrian Federal Ministry for Innovation, Mobility and Infrastructure (BMIMI), Austrian Federal Ministry for Economy, Energy and Tourism (BMWET), the Province of Styria (Dept. 12) and the Styrian Business Promotion Agency (SFG). The Austrian Research Promotion Agency (FFG) has been authorised for the programme management. Authors furthermore express thanks to the supporting industrial partner AVL List GmbH.

  \centering
  \includegraphics[scale=0.18]{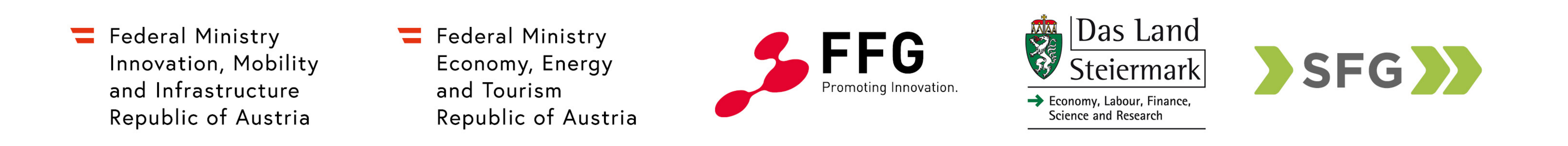} 

\end{document}